\documentclass[preprint,3p,number,sort&compress]{elsarticle}

\journal{Communications in Nonlinear Science and Numerical Simulation}

\usepackage{url}
\usepackage{ifxetex}
\ifxetex
\usepackage{polyglossia}
\setdefaultlanguage{english}
\else
\usepackage[utf8]{inputenc}
\usepackage[T1]{fontenc}
\usepackage[english]{babel}
\fi

\usepackage{amsfonts}
\usepackage{mathtools}
\usepackage{amssymb}

\usepackage{graphicx}

\usepackage{hyperref}
\hypersetup{colorlinks=true}

\newcommand{\onefig}[1]{\centering{\includegraphics[width=0.99\columnwidth]{#1}}}

\usepackage{bm}
\newcommand{\mat}[1]{\bm{\mathrm{#1}}}

\newcommand{\fpropag}{\mat{\mathcal{F}}}
\newcommand{\gpropag}{\mat{\mathcal{G}}}
\newcommand{\clvmat}{\mat{\Gamma}}

\newcommand{\mydd}{\mathrm{d}}
\newcommand{\adjnt}{*}
\newcommand{\metric}{\mat{M}}
\newcommand{\hmatrix}{\mat{H}}
\newcommand{\supT}{\text{T}}

\newcommand{\subbkw}{\text{b}}
\newcommand{\bkwmat}{\mat Q_\subbkw}
\newcommand{\bkwpre}{\widetilde{\mat Q}_\subbkw}
\newcommand{\bkwr}{\mat R_\subbkw}
\newcommand{\bkwa}{\mat A_\subbkw}

\newcommand{\subfwd}{\text{f}}
\newcommand{\fwdmat}{\mat Q_\subfwd}
\newcommand{\fwdpre}{\widetilde{\mat Q}_\subfwd}
\newcommand{\fwdr}{\mat R_\subfwd}
\newcommand{\fwda}{\mat A_\subfwd}

\newcommand{\const}{\mathrm{const}}
\newcommand{\diagmat}{\mathop{\mathrm{diag}}}

\newcommand{\fbas}{F}
\newcommand{\xbas}{X}
\newcommand{\ybas}{Y}
\newcommand{\ubas}{U}

\newcommand{\xvar}{x}
\newcommand{\yvar}{y}

\newcommand{\uvar}{u}
\newcommand{\vvarr}{v}
\newcommand{\vvar}[1]{v^{(#1)}}

\newcommand{\phdim}{N}
\newcommand{\nang}{K}

\newcommand{\locdim}{\nu}

\newcommand{\jacob}{\mat{J}}
\newcommand{\numjac}{\mat{\mathcal{J}}}
\newcommand{\delnum}{d}

\newcommand{\ntau}{k}

\begin{document}

\begin{frontmatter}
  \title{Numerical test for hyperbolicity in chaotic systems
    with multiple time delays}

  \author[sstu]{Pavel V. Kuptsov\corref{cor1}}%
  \cortext[cor1]{Corresponding author}
  \ead{p.kuptsov@rambler.ru}

  \author[udsu,ire]{Sergey P. Kuznetsov}%
  \ead{spkuz@yandex.ru}

  \address[sstu]{Institute of electronics and mechanical engineering, Yuri
  Gagarin State Technical University of Saratov, Politekhnicheskaya
  77, Saratov 410054, Russia}%

  \address[udsu]{Institute of Mathematics, Information Technologies and
    Physics, Udmurt State University, Universitetskaya 1, Izhevsk
    426034, Russia}%

  \address[ire]{Kotel’nikov’s Institute of Radio-Engineering and
    Electronics of RAS, Saratov Branch, Zelenaya 38, Saratov 410019,
    Russia}%

  \begin{abstract}
    We develop an extension of the fast method of angles for
    hyperbolicity verification in chaotic systems with an arbitrary
    number of time-delay feedback loops. The adopted method is based
    on the theory of covariant Lyapunov vectors and provides an
    efficient algorithm applicable for systems with high-dimensional
    phase space.  Three particular examples of time-delay systems are
    analyzed and in all cases the expected hyperbolicity is confirmed.
  \end{abstract}

  \begin{keyword}
    hyperbolic chaos \sep
    hyperbolicity test \sep
    fast method of angles \sep
    delay differential equations
  \end{keyword}

\end{frontmatter}

\section{Introduction}

Hyperbolic theory~\cite{Smale67,Anosov95,KatHas95} studies invariant
sets in phase space of dynamical systems, including those with chaotic
dynamics, composed exclusively of saddle trajectories. For all points
on such a trajectory, in the space of small perturbations (tangent
space), one can define a subspace of vectors, which exponentially
decrease in norm under the forward time evolution, and a subspace of
vectors, which exponentially decrease under the backward time
evolution. In flow systems, in addition, there is a one-dimensional
neutral subspace of perturbations along a trajectory that neither
increase nor decrease on average. An arbitrary vector of small
perturbation is a linear combination of vectors belonging to the
indicated subspaces. A set of states that approach a given trajectory
during time evolution is called the contracting (or stable) manifold
of this trajectory. Similarly, the expanding (unstable) manifold is a
set of states tending to the reference trajectory under the backward
time evolution. Tangencies of stable and unstable manifolds should be
excluded; only intersections at nonzero angles are admitted.

Hyperbolic chaos plays a special role among other types of chaotic
dynamics. Systems of this type, like, for example, the Smale-Williams
solenoid, manifest deterministic chaos justified in rigorous
mathematical sense. They demonstrate strong and structurally stable
stochastic properties. They are insensitive to variation of functions
and parameters in the dynamical equations, to noises, interferences
etc. Moreover, hyperbolic chaotic dynamics in such systems allow a
detailed mathematical analysis~\cite{Smale67,Anosov95,KatHas95}.

Due to their great potential importance for applications, structurally
stable chaotic systems with hyperbolic attractors obviously have to be
a subject of priority interest, like rough systems with regular
dynamics in the classic theory of
oscillations~\cite{AndrPontr,AndrKhaikVitt}. However, many years the
hyperbolic attractors were commonly regarded only as purified abstract
mathematical images of chaos rather than something intrinsic to real
world systems. A certain progress in this field has been achieved
recently when many examples of physically realizable systems with
hyperbolic attractor have been purposefully constructed using a
toolbox of physics (oscillators, particles, fields, interactions,
feedback loops, etc.)~\cite{HyperBook12,KuzUFN11}.

Physical and technical devices employed in the offered systems surely
are not well suited for rigorous mathematical analysis. However,
strong evidence of the hyperbolicity is significant to exploit
properly relevant theoretical results. In this situation numerical
verification of hyperbolicity of chaos becomes an essential ingredient
of the studies.

Systems with time-delay feedback combine relative simplicity of
implementation, almost like low dimensional systems modeled by
ordinary differential equations, and rich complexity of dynamics,
comparable with infinite dimensional systems associated with partial
differential equations. Examples of such systems are wide-spread in
electronics, laser physics, acoustics and other
fields~\cite{VihLaf13}. Their mathematical description is based on
differential equations with
delays~\cite{BellCook63,Myshkis72,ElsNor73}. A number of time-delay
systems with chaotic dynamics was
explored~\cite{DorGram87,Chrost83,Lepri94,Mackey77,
  Farmer82,GrassProc84,KuzPonom08,Autonom10,
  KuzPik08,Baranov10,KuzKuz13}, and several examples were suggested as
realizable devices for generation of rough hyperbolic
chaos~\cite{KuzPonom08,Autonom10,
  KuzPik08,Baranov10,KuzKuz13,Arzh14}. However, no direct verification
of the hyperbolicity was performed for them.

There are two different approaches to numerical test of
hyperbolicity. One of them, the method of cones, is based on a
mathematical theorem on expanding and contracting
cones~\cite{Sinai79,KatHas95}. It has been adopted for computer
verification and applied for some low-dimensional
systems~\cite{KuzSat07,KuzKuzSat10,Wilcz10}.

The second approach, the method of angles, directly employs the
definition of hyperbolic attractor: its orbits are of saddle type, and
their expanding and contracting manifolds do not have tangencies but
can only intersect transversally. The method involves a computation of
angles between the manifolds along trajectories. In the case of
hyperbolicity, the distributions of these angles are distant from
zero. This method was used initially in Ref.~\cite{LaiGeb93}, and in
Ref.~\cite{FastHyp12} its fast and economical reformulation was
suggested. This approach may be regarded as an extension of Lyapunov
analyses, well-established and applied successively not only for
low-dimensional systems but for spatiotemporal systems
too~\cite{Hirata99,Anishch00,Kuznetsov05,KuzSelez06,KupKuz09,Kuznetsov15,
  KuzKrug16,Kuz16}.

In our recent brief report~\cite{KupKuz16} we have adopted the fast
method of angles~\cite{FastHyp12} to perform the hyperbolicity test
for systems with a single time delay. In the present paper we extend
this method for
systems with arbitrary number of time-delay feedbacks. Three
particular examples considered in
Refs.~\cite{Baranov10,KuzPik08,Arzh14} are analyzed, and in all cases
the expected hyperbolicity is confirmed.

The paper is organized as follows. In Sec.~\ref{sec:theory} we briefly
review the theory laying behind the fast method of angles. Required
for this method adjoint variational time-delay equation is derived in
Sec.~\ref{sec:analyt_adj}, and in Sec~\ref{sec:numeric_adj} its
numerical approximation is considered. In Sec.~\ref{sec:num_scheme} we
discuss a numerical method for solving delay differential equations
that is applied in Sec.~\ref{sec:examples} for hyperbolicity tests of
particular time-delay systems. Finally in Sec.~\ref{sec:concl} the
results of the paper are summarized.

\section{\label{sec:theory}Numerical verification of hyperbolic chaos.
  Fast numerical algorithm for the method of angles}

In this section we review the theoretical background of the fast
method of angles. For more details see
Refs.~\cite{CLV2012,FastHyp12}. The method is based on the concept of
covariant Lyapunov vectors
(CLVs)~\cite{GinCLV,WolfCLV,CLV2012,PikPol16}. The adjective
``covariant'' means that these vectors are in one-to-one
correspondence with Lyapunov exponents and evolve in such way that the
$i$th CLV at time $t_1$ is mapped to the $i$th CLV at $t_2$. In
average they grow or decay exponentially, each with the corresponding
Lyapunov exponent.

CLVs form a natural non-orthogonal basis for the subspaces tangent to
expanding, neutral and contracting manifolds. Analysis of these
vectors can reveal presence or violation of the hyperbolicity. As
follows from the definition~\cite{Smale67,Anosov95,KatHas95}, a chaotic
system is hyperbolic when the manifolds never have tangencies. So, the
idea is to compute angles between CLVs related to the positive, zero
and negative Lyapunov exponents. The system is hyperbolic if these
angles never vanish.

There are various methods for computations of CLVs. The fast method of
angles employs the one suggested in Ref.~\cite{CLV2012}. Its idea goes
back to the work~\cite{WolfCLV} with later essential supplement
form~\cite{PazoSzendro08}. The other method for CLVs can be found in
Ref.~\cite{GinCLV}. See also a book~\cite{PikPol16} for an extended
review.

CLVs can be computed both for continues and discrete time systems.
Consider a system represented by an ordinary differential equation:
\begin{equation}
  \label{eq:com_bas_sys}
  \dot\xbas=\fbas(\xbas,t),
\end{equation}
where $\xbas\equiv\xbas(t)\in\mathbb{R}^\phdim$ is $\phdim$
dimensional state vector, and $\fbas$ is a nonlinear function of
$\xbas$ and, for non-autonomous systems, of $t$. Infinitely small or
so called tangent perturbations to trajectories of the
system~\eqref{eq:com_bas_sys} obey variational equation
\begin{equation}
  \label{eq:com_var_sys}
  \dot\xvar=\jacob(t) \xvar,
\end{equation}
where $\xvar\equiv\xvar(t)\in\mathbb{R}^\phdim$ is a tangent vector
and $\jacob(t)\equiv\jacob(X,t)\in \mathbb{R}^{\phdim\times\phdim}$ is
the Jacobian matrix, i.e., the matrix of derivatives of $\fbas$ with
respect to $\xbas$. Its time dependence can be implicit via $X(t)$ and
explicit in the non-autonomous case.

Deriving the variational equation~\eqref{eq:com_var_sys} we do not
automatically find out the way of computation of tangent vector
norms. In other words the basis for these vectors remain unknown. We
have to define it introducing a metric tensor $\metric$. Usually the
identity matrix is taken, but we will consider a more general case of
real symmetric positive definite metrics $\metric$.  Once such metric
is selected, an inner product of two arbitrary vectors $a$ and $b$ is
defined as~\cite{Horn2012,Hogben2013}
\begin{equation}
  \label{eq:inner_prod}
  \langle a,b\rangle =a^\supT \metric b,
\end{equation}
where ``$\supT$'' stands for
transposition. In turn, the inner product
allows evaluating the vector norms:
\begin{equation}
  \label{eq:def_norm}
  ||a||^2=\langle a,a\rangle =a^\supT \metric a.
\end{equation}

Vectors of the basis itself are represented as $(1,0,0,\ldots)^\supT$,
$(0,1,0,\ldots)^\supT$ and so on. It can be easily checked that for a
generic metric $\metric$ these vectors are neither orthogonal nor
normalized. The following transformation provides the
orthonormalization:
\begin{equation}
  \label{eq:to_orthonorm_basis}
  a'=\hmatrix a,
\end{equation}
where $\hmatrix$ is such that $\hmatrix^\supT\hmatrix=\metric$. The
matrix $\hmatrix$ always exists since $\metric$ is assumed to be
positive definite and symmetric.  The metric $\metric$ under this
transformation changes into an identity matrix so that the inner
product takes the form of the standard dot product:
$a^\supT \metric b=(a')^\supT b'$.

Lyapunov exponents as well as CLVs do not depend on the metric
choice. That is why usually the simplest case is considered, i.e., the
identity matrix is taken as the metric and the standard dot product is
used. However, as we will see in the subsequent sections time-delay
systems need more accurate approach.

Evolution of a tangent vector from time $t_1$ to time $t_2$ can be
expressed as action of a linear operator $\fpropag(t_1,t_2)$ called
propagator:
\begin{equation}
  \xvar(t_2)=\fpropag(t_1,t_2)\xvar(t_1).
\end{equation}
For the continues time system~\eqref{eq:com_var_sys} the propagator
$\fpropag(t_1,t_2)$ is related with the Jacobian matrix $\jacob(t)$
through the Magnus expansion~\cite{CLV2012}. In actual numerical
computations it merely means that we solve Eq.~\eqref{eq:com_var_sys}
from $t_1$ to $t_2$ to find a result of the propagator action.

To compute CLVs and find corresponding angles we employ the standard
algorithm for Lyapunov exponents~\cite{Benettin80,Shimada79}. First, a
required number $\nang$ of tangent vectors are initialized and
written as columns of a matrix $\bkwmat(t_1)$. Then the propagator is
applied to obtain $\bkwpre(t_2)$, that actually means solving the
basic system together with $\nang$ copies of the variational
equations.

Due to action of the propagator, any arbitrarily chosen vector tends to line up
along the most expanding direction growing exponentially with the
rate equal to the largest Lyapunov exponent. Similarly, any
non-colinear pair of vectors tends to the most expanding plane, and the spanned area grows
exponentially with a rate determined by the sum of two first Lyapunov
exponents. Any three vectors approach the most expanding
three-dimensional volume with the growth rate equal to the sum of
three first Lyapunov exponents and so on.

Each previous alignment shades the next one: without a special
treatment we will not see the most expanding plane since the expanding
direction will absorb all its vectors. But it becomes available when
an orthogonalization procedure of vector-columns of $\bkwpre(t_2)$ is
performed in the course of the computations.
The first vector preserves its direction; the second one is
rotated up to the orthogonal position always staying on the plane
spanned by its initial direction and the first vector; the third
vector is rotated to become orthogonal to the first two but strictly
within the space spanned by the first three vectors and so on. The
rotations are accompanied by normalization of the vector lengths. This
procedure is known as Gram-Schmidt orthogonalization or QR
factorization~\cite{GolubLoan,Hogben2013}. The letter means
representation of a matrix as a product an orthogonal matrix $\mat Q$
and an upper triangular matrix $\mat R$.

Altogether, the iteration step includes the evolution from $t_1$ to
$t_2$ and the orthogonalization:
\begin{gather}
  \label{eq:forward_lyap_step}
  \fpropag(t_1,t_2)\bkwmat(t_1)=\bkwpre(t_2),\\
  \label{eq:forward_qr}
  \bkwpre(t_2)=\bkwmat(t_2)\bkwr(t_2).
\end{gather}
Computed in this way orthogonal matrix $\bkwmat(t_2)$ is used for the
next step of the algorithm.

After skipping transients, we can start to accumulate absolute values
of logarithms of diagonal elements of $\bkwr(t)$. Averaged over the
iteration time they become Lyapunov exponents. CLVs computation
requires $\bkwmat(t)$ instead. After the transient, the columns of
this orthogonal matrix become the backward Lyapunov vectors. This name
emphasizes the fact that these vectors arrive at time $t$ after long
evolution form far past. Also they are known as Gram-Schmidt
vectors. Being norm-dependent, they nevertheless contain an essential
information about the tangent structure of the trajectory
manifolds. If CLVs are gathered as columns of a matrix $\clvmat(t)$
they can be represented via backward Lyapunov vectors~\cite{CLV2012}:
\begin{equation}
  \label{eq:clv_via_bkw}
  \clvmat(t)=\bkwmat(t)\bkwa(t),
\end{equation}
where $\bkwa(t)$ is an upper triangular matrix. It means that the first
CLV coincides with the first backward Lyapunov vector, the second one
belongs to a plane spanned by the first two backward vectors, the third one
belongs to a three-dimensional space of the first three backward
vectors and so on.

If $\nang$ is dimension of the expanding tangent subspace, i.e., the
number of positive Lyapunov exponents, from iterations of
\eqref{eq:forward_lyap_step},~\eqref{eq:forward_qr} with $\nang$
vectors we obtain the basis for this subspace.  To carry out the
hyperbolicity check we also need the basis for the remaining
contracting subspace whose dimension is $\phdim-\nang$. At the moment
we ignore possible existence of the neutral subspace associated with
zero Lyapunov exponents.

The absent basis can be computed in the course of iterations of tangent vectors backward in time. The straightforward idea is to perform
steps analogous to~\eqref{eq:forward_lyap_step} and
\eqref{eq:forward_qr} but with $\fpropag^{-1}$. In other words, we can
integrate $\phdim-\nang$ copies of the variational equation with
negative time step. This approach was used in early papers when the
theory of CLVs was not well developed, and, in particular, no
efficient algorithms for their computations were known. It worked well
for low dimensional systems, but if $\phdim$ was high it became
inapplicable due to very high consumption of computational resources.

The preferred way is to perform backward steps with an adjoint
propagator $\fpropag^\adjnt$ satisfying the identity
$\langle \fpropag^\adjnt a,b\rangle\equiv \langle a,\fpropag
b\rangle$~\cite{Horn2012,Hogben2013}, where $a$ and $b$ are arbitrary
vectors. Taking~\eqref{eq:inner_prod} into account we
obtain\footnote{Notice that in Ref.~\cite{CLV2012} the adjoint
  propagator is introduced as
  $\gpropag=\left[\fpropag^\adjnt\right]^{-1}$.}:
\begin{equation}
  \label{eq:adjoint_propag}
  \fpropag^\adjnt=\metric^{-1}\fpropag^\supT\metric.
\end{equation}
The implementation of this propagator includes solving the adjoint
variational equation
\begin{equation}
  \label{eq:com_adj_sys}
  \dot\yvar=-\jacob^\adjnt(t)\yvar
\end{equation}
backward in time, where
$\jacob^\adjnt(t)=\metric^{-1}[\jacob(t)]^\supT\metric$ is the adjoint
Jacobian matrix. As discussed in Refs.~\cite{AGuide,CLV2012} the
iterations with $\fpropag^{-1}$ and $\fpropag^\adjnt$ generate
identical sets of vectors, but since the inverted propagator
$\fpropag^{-1}$ has the reciprocal singular values with respect to
$\fpropag^\adjnt$, in its iterations the smallest Lyapunov exponent
dominates so that the resulting vectors come in the reverse order. The
use of the adjoint propagator instead of the inverted one is the key
point of the fast method of angles~\cite{FastHyp12}.

Consider the following backward time iterations with
$\fpropag^\adjnt$:
\begin{gather}
  \label{eq:bkwadj_lyap_step}
  \fpropag(t_1,t_2)^\adjnt\fwdmat(t_2)=\fwdpre(t_1),\\
  \label{eq:bkwadj_qr}
  \fwdpre(t_1)=\fwdmat(t_1)\fwdr(t_1).
\end{gather}
Here an orthogonal matrix $\fwdmat(t)$ contains $\nang$ columns in
exactly the same way as $\bkwmat(t)$ in
Eqs.~\eqref{eq:forward_lyap_step}, \eqref{eq:forward_qr}.  After the
transient, the columns become the forward Lyapunov vectors, i.e., the
vectors arriving at $t$ after long iteration initiated in far
future. The full set of $\phdim$ these vectors is related with CLVs as
follows, cf. Eq.~\eqref{eq:clv_via_bkw}:
\begin{equation}
  \label{eq:clv_via_fwd}
  \clvmat(t)=\fwdmat(t)\fwda(t),
\end{equation}
where $\fwda(t)$ is a lower triangular matrix~\cite{CLV2012}. It means
that the first $\nang$ forward vectors form an orthogonal complement
for the sought subspace of the last $\phdim-\nang$ CLVs. The important
point is that typically the complement has much lower dimension, and
the corresponding computational routine, including
iterations~\eqref{eq:bkwadj_lyap_step},~\eqref{eq:bkwadj_qr} with
$\nang$ columns of $\bkwmat(t)$ is more economical and fast than straightforward
iterations of $\phdim-\nang$ vectors with the inverted propagator
$\fpropag^{-1}$.

Altogether, the first $\nang$ backward Lyapunov vectors span the
subspace holding the first $\nang$ CLVs, and the first $\nang$ forward
Lyapunov vectors provide the basis for the orthogonal complement to
the subspace holding the last $\phdim-\nang$ CLVs. To check the
tangency we need to find principal angles between these two
subspaces~\cite{GolubLoan}. Given their orthogonal bases $\bkwmat(t)$
and $\fwdmat(t)$, respectively, cosines of the principal angles are
found as singular values of $\nang\times\nang$ matrix of pairwise
inner products of the bases vectors:
\begin{equation}
  \label{eq:matrix_p}
  \mat P(t)=[\fwdmat(t)]^\adjnt\bkwmat(t).
\end{equation}
The second subspace is the orthogonal complement to the subspace of
interest. It means that the tangency is signaled by the largest
principal angle that corresponds to the smallest singular value. Once
the matrix $\mat P$ has been computed, we gain access to a series of
$\nang$ angles. Taking its top left square submatrix
$\mat P[1:i,1:i]$, where $i=1,2,\ldots,\nang$, and finding the smallest
corresponding singular value $\sigma_i$, we can compute the angle
between the $i$ dimensional subspace of the first CLVs and $\phdim-i$
dimensional subspace of the remaining CLVs as follows:
\begin{equation}
  \label{eq:the_angle}
  \theta_i=\pi/2-\arccos\sigma_i.
\end{equation}

The smallest singular value $\sigma_i$ as well as the angle $\theta_i$
vanish when a tangency between the corresponding subspaces occurs.
Because trajectories with the exact tangencies are rather untypical,
in actual computations we register a tangency between subspaces if the
corresponding angle can be arbitrarily small.

Since the angle computation involves norm dependent forward and
backward Lyapunov vectors, particular values of $\theta_i$ depend on
the norm~\eqref{eq:def_norm}, i.e., on the choice of the metric $\metric$. However, the related topological properties, i.e., their
vanishing or non-vanishing, are norm independent.

Typical examples of application of the described approach are the
following. When a chaotic system has a single positive Lyapunov exponent
and zero exponents are absent, e.g. for a discrete-time system, or for
a periodically driven non-autonomous system, we have $\nang=1$ and
need to perform forward
steps~\eqref{eq:forward_lyap_step},~\eqref{eq:forward_qr} and backward
steps \eqref{eq:bkwadj_lyap_step},~\eqref{eq:bkwadj_qr} monitoring one
backward and one forward Lyapunov vectors, respectively. The matrix
$\mat P$~\eqref{eq:matrix_p} is reduced to an inner product of these
vectors that is substituted as $\sigma_1$ to Eq.~\eqref{eq:the_angle}
to compute $\theta_1$. The hyperbolicity is confirmed when the
distribution of this angle is clearly separated from zero.

Another example is an autonomous continues time chaotic system with
one positive and one zero Lyapunov exponents. In this case $\nang=2$
so that forward~\eqref{eq:forward_lyap_step},~\eqref{eq:forward_qr}
and backward~\eqref{eq:bkwadj_lyap_step},~\eqref{eq:bkwadj_qr}
iterations are performed with two vectors and two angles are
computed. For $\theta_1$ we again take the top left element of
$\mat P$ as $\sigma_1$, and for $\theta_2$ the smallest singular value
$\sigma_2$ of $2\times 2$ matrix $\mat P$ is found. The hyperbolicity
is confirmed if the system corresponds to the Anosov
flow~\cite{Anosov95,KatHas95}: its expanding, neutral and contracting
subspaces never clash. This is the case when the distributions both
for $\theta_1$ and for $\theta_2$ are well separated from the origin.

\section{\label{sec:analyt_adj}Adjoint variational equation
  for a system with multiple time delays}

Consider a system with $\delnum$ delays:
\begin{equation}
  \label{eq:main_eq}
  \dot\xbas=\fbas[t,\xbas(t),\xbas(t-\tau_1),\ldots,\xbas(t-\tau_\delnum)].
\end{equation}
Here $\xbas(t)\in \mathbb{R}^\locdim$ is a vector variable of finite
dimension $\locdim$.
Delay times $\tau_i$ are
assumed to be labeled in the ascending order, so that $\tau_1$ is the
shortest one and $\tau_\delnum$ is the longest one.

The full dimension of the phase space is infinite: one needs to
consider a trajectory segment of duration $\tau_\delnum$ (a continuum
of data) to determine each new infinitesimal time step.

The corresponding variational equation reads
\begin{equation}
  \label{eq:main_var_eq}
  \dot\xvar=\jacob_0(t)\xvar(t)+\sum_{i=1}^\delnum \jacob_i(t)\xvar(t-\tau_i),
\end{equation}
where $\jacob_0(t)$ and $\jacob_{i\geq 1}(t)$ are the derivative
matrices composed of partial derivatives of $\fbas$ over components of
$\xbas$ and $\xbas_{\tau_i}$, respectively.

To apply the fast method of angles to system~\eqref{eq:main_eq} we
need to the adjoint variational equation
to~\eqref{eq:main_var_eq}. Note that arbitrary solutions $\xvar(t)$
and $\yvar(t)$ to the variational and the adjoint
equations~\eqref{eq:com_var_sys} and \eqref{eq:com_adj_sys},
respectively, must fulfill the identity
\begin{equation}
  \label{eq:adj_cond}
  \frac{\mydd}{\mydd t} \left\langle \xvar(t),\yvar(t)\right\rangle
  \equiv 0,
\end{equation}
as can be verified by direct substitution. We will find the adjoint
equation for Eq.~\eqref{eq:main_var_eq} requiring fulfillment of the
analogous identity.

In an actual physical implementation, the system~\eqref{eq:main_eq} may be
thought as endowed with a delay line providing the retarded variables
$\xbas_{\tau_i}$. A natural way to take it into account explicitly is
to introduce a wave system with a delta-function source at the origin:
\begin{equation}
  \label{eq:delay_expl_for_orig_sys0}
  \ubas_t+\ubas_\xi=\delta(\xi)\xbas(t).
\end{equation}
Here $\ubas\equiv \ubas(t,\xi)$ is the delay line variable depending on
the coordinate $\xi$ and time $t$. The subscripts
$t$ and $\xi$ stand for the corresponding partial derivatives. The
solution to Eq.~\eqref{eq:delay_expl_for_orig_sys0} is a wave
propagating in the positive direction:
\begin{equation}
  \label{eq:delay_expl_sol0}
  \ubas(t,\xi)=\xbas(t-\xi).
\end{equation}
Now, the main equation~\eqref{eq:main_eq} can be rewritten as
\begin{equation}
  \label{eq:main_eq1}
  \dot\xbas=\fbas[t,\xbas(t),\ubas(t, \tau_1),\ldots,\ubas(t, \tau_\delnum)].
\end{equation}
Respectively, the variational equation takes the form
\begin{equation}
  \label{eq:delay_expl_for_orig_sys}
  \uvar_t+\uvar_\xi=\delta(\xi)\xvar(t),
\end{equation}
\begin{equation}
  \label{eq:delay_expl_sol}
  \uvar(t,\xi)=\xvar(t-\xi),
\end{equation}
and
\begin{equation}
  \label{eq:main_var_eq_with_delay_expl}
  \dot\xvar=\jacob_0(t)\xvar(t)+\sum_{i=1}^\delnum \jacob_i(t)\uvar(t,\tau_i),
\end{equation}
where $\jacob_0(t)$ and $\jacob_{i\geq 1}(t)$ are the same matrices as
used in~\eqref{eq:main_var_eq}.

A state vector for Eq.~\eqref{eq:main_var_eq_with_delay_expl} has a
mixed discrete-continues form, $\bar{\xvar}=(\xvar,u)$. The inner
product for two such vectors $\bar{\xvar}=(\xvar,\uvar)$ and
$\bar{\yvar}=(\yvar,\vvarr)$ can be introduced as
\begin{equation}
  \label{eq:delay_inner_def}
  \langle\bar{\xvar},\bar{\yvar}\rangle=
  \xvar^\supT(t)\yvar(t)+
  \int_0^{\tau_\delnum}\uvar^\supT(t,\xi)\vvarr(t,\xi) \,\mydd\xi.
\end{equation}

Now we will construct the adjoint variational equation requiring
fulfillment of the identity
$\mydd\left\langle \bar{\xvar},\bar{\yvar}\right\rangle /\mydd t\equiv
0$ with respect to the inner product~\eqref{eq:delay_inner_def}. The
desirable equation reads
\begin{equation}
  \label{eq:adj_sys_with_delay_expl}
  \dot\yvar=-\jacob_0^\supT(t)\yvar(t)-\vvar{1}(t,0).
\end{equation}
Here $\vvar{1}(t,\xi)$ is the first segment of a compound delay line
including the following $\delnum$ parts:
\begin{equation}
  \label{eq:coumpund_delay_line}
  \begin{aligned}
    \vvar{i}_t+\vvar{i}_\xi&=\delta(\xi+\tau_i)
    [\jacob_i^\supT(t)\yvar(t)+\vvar{i+1}(t,\xi)],\\
    \vvar{\delnum}_t+\vvar{\delnum}_\xi&=\delta(\xi+\tau_\delnum)
    \jacob_\delnum^\supT(t)\yvar(t),
  \end{aligned}
\end{equation}
where $i=1,\ldots,\delnum-1$.  Solution to $n$th segment reads:
\begin{equation}
  \label{eq:coumpund_delay_line_sol}
  \vvar{n}(t,\xi)=\sum_{i=n}^\delnum \jacob_i^\supT(t-\xi+\tau_i)y(t-\xi+\tau_i),
\end{equation}
where $n=1,\ldots,\delnum$.

The delay line~\eqref{eq:delay_expl_for_orig_sys} contains a source at
the origin where the signal $\xvar(t)$ is injected. The retarded
signals are read at points $\xi=\tau_i$ and returned back to the
system. The delay line~\eqref{eq:coumpund_delay_line} for the adjoint
system~\eqref{eq:adj_sys_with_delay_expl} is a chain of $\delnum$
segments coupled via a kind of sinks at their right boundaries. The
signal $\yvar(t)$ is injected through every sink being multiplied by
the corresponding Jacobian matrix. The advanced wave solution of the
first chain segment is read at the origin and returned back to the
system.

To confirm the correctness of the adjoint
equation~\eqref{eq:adj_sys_with_delay_expl} we can expand the identity
$\mydd\left\langle \bar{\xvar},\bar{\yvar}\right\rangle /\mydd t\equiv
0$ using the inner product~\eqref{eq:delay_inner_def} as
\begin{equation}
  \label{eq:inner_prod_zero_check}
  \frac{\mydd}{\mydd t}
  \left[ \xvar^\supT(t)\yvar(t)+\sum_{i=1}^{\delnum}
    \int_{\tau_{i-1}}^{\tau_i}
    \uvar^\supT(t,\xi)\vvar{i}(t,\xi)\mydd\xi \right]\equiv 0,
\end{equation}
where $\tau_0=0$. Verification of \eqref{eq:inner_prod_zero_check} is
rather straightforward taking into account the equality
$(\mydd/\mydd t) \int_a^bf(t-\xi)\mydd \xi=f(t-a)-f(t-b)$.

Substituting Eq.~\eqref{eq:coumpund_delay_line_sol}
to~\eqref{eq:adj_sys_with_delay_expl}, we finally obtain the adjoint
variational equation as a differential equation with deviating
argument:
\begin{equation}
  \label{eq:main_adj_eq}
  \dot\yvar=-\jacob_0^\supT(t)\yvar(t)-
  \sum_{i=1}^\delnum \jacob_{i}^\supT(t+\tau_i) \yvar(t+\tau_i).
\end{equation}

In the theory of differential equations with deviating arguments
Eq.~\eqref{eq:main_adj_eq} belongs to the class of equations of
leading or advanced type~\cite{BellCook63,Myshkis72,ElsNor73}. They
are regarded as poorly defined with respect to the existence of
solutions to initial value problems. In the context of our study,
however, we will solve such equations in backward time only, so that
they behave in a good way like the equations of retarded type in
forward time.

\section{\label{sec:numeric_adj}Numerical approximation of the adjoint
  variational equations}

In the previous section we have shown that the adjoint companion for the
time-delay variational Eq.~\eqref{eq:main_var_eq} is
Eq.~\eqref{eq:main_adj_eq} providing that the inner product is defined
by Eq.~\eqref{eq:delay_inner_def}. In this section we will show that
Eq.~\eqref{eq:main_adj_eq} agrees with the adjoint numerical Jacobian,
i.e., performing numerical simulations we can either solve the adjoint
equation~\eqref{eq:main_adj_eq} directly or find the numerical
Jacobian matrix for Eq.~\eqref{eq:main_var_eq} and then compute its
adjoint form. For the sake of simplicity, we consider here the Euler numerical scheme of the first order.

Let $h>0$ be a time step and let us assume that $\ntau_i=\tau_i/h$,
where all $\ntau_i$, $i=1,\ldots,\delnum$, are integers. Since
$\tau_i$ are ordered, $\ntau_i$ are ordered too:
$\ntau_1<\ldots<\ntau_\delnum$.  Also we set $t_n=nh$, $\xi_i=ih$, and
$\uvar_{n,i}\equiv\uvar(t_n,\xi_i)$, $\xvar_n\equiv
\xvar(t_n)$. Accepting these assumptions we obtain the Euler numerical
approximation for the variational
Eq.~\eqref{eq:main_var_eq_with_delay_expl} as follows:
\begin{equation}
  \label{eq:euler_main}
  \frac{\xvar_{n+1}-\xvar_n}{h}=\jacob_0(t_n)\xvar_n
  +\sum_{i=1}^\delnum \jacob_i(t_n) \uvar_{n,\ntau_i},
\end{equation}
where $\uvar_{n,i}$ is a solution of a discrete form of
Eq.~\eqref{eq:delay_expl_for_orig_sys} for the delay line:
\begin{equation}
  \label{eq:euler_delay_line}
  \uvar_{n,0}=\xvar_{n},\; \uvar_{n+1,i}=\uvar_{n,i-1}.
\end{equation}

Equations~\eqref{eq:euler_main} and \eqref{eq:euler_delay_line} admit
recasting in a matrix form
$\bar\xvar_{n+1}=\numjac(t_n)\bar\xvar_{n}$, where $\numjac(t_n)$ is a
numerical Jacobian matrix playing a role of the propagator applicable
for the forward time
iterations~\eqref{eq:forward_lyap_step},~\eqref{eq:forward_qr}, see
Sec.~\ref{sec:theory}. For example at $\delnum=2$, $\ntau_1=2$, and
$\ntau_2=4$ the numerical Jacobian matrix reads:
\begin{equation}
  \label{eq:example_numeric_jac}
  \numjac(t_n)=
  \begin{pmatrix}
    1+h\jacob_0(t_n) & 0 & h\jacob_1(t_n) & 0 & h \jacob_2(t_n) \\
    1 & 0 & 0 & 0 & 0 \\
    0 & 1 & 0 & 0 & 0 \\
    0 & 0 & 1 & 0 & 0 \\
    0 & 0 & 0 & 1 & 0
  \end{pmatrix}.
\end{equation}
In general, this is a block matrix
$(\ntau_\delnum+1)\times(\ntau_\delnum+1)$ that contains
$h\jacob_i(t_n)$ at sites $k_i$ of the first row, ones, i.e., identity
blocks, on the first subdiagonal and other elements are zeros.

Given $\numjac(t_n)$ we can find an explicit form of the adjoint
numerical Jacobian matrix $\numjac^\adjnt(t_n)$ applicable for the
backward iterations \eqref{eq:bkwadj_lyap_step}, \eqref{eq:bkwadj_qr}.
Since the adjoint variational equation~\eqref{eq:main_adj_eq} is
constructed with respect to the inner
product~\eqref{eq:delay_inner_def}, now we need to discretize it as
follows
\begin{equation}
  \label{eq:delay_inner_def_dis}
  \langle\bar\xvar_n,\bar\yvar_n\rangle=
  \bar\xvar_n^\supT \mat H^2\bar\yvar_n=
  \xvar^\supT_n\yvar_n+
  h\sum_{i=1}^{\ntau_\delnum} \uvar^\supT_{n,i}\vvarr_{n,i},
\end{equation}
where
$\bar\xvar_n=(\xvar_n,\uvar_{n,1},\ldots,\uvar_{n,\ntau_\delnum})$ and
$\bar\yvar_n=(\yvar_n,\vvarr_{n,1},\ldots,\vvarr_{n,\ntau_\delnum})$
are state vectors and a diagonal matrix
$\hmatrix^2\in
\mathbb{R}^{(\ntau_\delnum+1)\times(\ntau_\delnum+1)}$ plays the role
of a metric,
\begin{equation}
  \label{eq:the_metric}
  \metric=\hmatrix^2,\;
  \hmatrix=\diagmat(1,\sqrt{h},\ldots ,\sqrt{h}).
\end{equation}
Taking into account the definition of the adjoint
propagator~\eqref{eq:adjoint_propag} we obtain the adjoint numerical
Jacobian matrix:
\begin{equation}
  \label{eq:adjoint_numeric_jac}
  \numjac^\adjnt(t_n)=\hmatrix^{-2}\numjac^\supT(t_n) \hmatrix^2.
\end{equation}
For the matrix~\eqref{eq:example_numeric_jac} the
transformation~\eqref{eq:adjoint_numeric_jac} results in
\begin{equation}
  \label{eq:example_numeric_adj}
  \numjac^\adjnt(t_n)=
  \begin{pmatrix}
    1+h\jacob_0^\supT & h & 0 & 0 & 0\\
    0 & 0 & 1 & 0 & 0\\
    \jacob_1^\supT & 0 & 0 & 1 & 0 \\
    0 & 0 & 0 & 0 & 1 \\
    \jacob_2^\supT & 0 & 0 & 0 & 0
  \end{pmatrix}.
\end{equation}

When two manifolds of a trajectory have a tangency, i.e., the
corresponding angle vanishes, this property is preserved under the
metric change. It means that we could also detect this situation
taking the unit matrix as a metric and using the standard dot product
instead of Eq.~\eqref{eq:delay_inner_def_dis}. However the angles
computed in this way would depend on the discretization step $h$. The
inner product \eqref{eq:delay_inner_def_dis} provides a correct
asymptotic behavior of the angles as $h\to 0$ when the numerical
scheme converges to the original differential equations. This is the
case because Eq.~\eqref{eq:adjoint_numeric_jac} with the
metric~\eqref{eq:the_metric} produces the adjoint numerical Jacobian
matrix $\numjac^\adjnt(t_n)$ that corresponds to the Euler
discretization of the adjoint variational equation. Let
$y_n\equiv y(t_n)$, $\vvarr_{n,i}\equiv\vvarr(t_n,\xi_i)$,
$t_n=t_0-nh$, $\xi_n=ih$, and $h>0$. In these terms the equation for
backward time iterations is as follows:
$\bar\yvar_{n+1}=\numjac^\adjnt(t_n)\bar\yvar_n$, where
$\bar\yvar_n=(\yvar_n,\vvarr_{n,1},\ldots,\vvarr_{n,\ntau_\delnum})$. The
Euler discretization of the adjoint variational
Eq.~\eqref{eq:adj_sys_with_delay_expl} for backward time solution
exactly corresponds to this map. It can be illustrated using the
matrix~\eqref{eq:example_numeric_adj} whose iteration step reads:
\begin{equation}
  \begin{aligned}
    \frac{\yvar_{n+1}-\yvar_n}{-h}&=-\jacob_0^\supT\yvar_n-\vvarr_{n,1},\\
    \vvarr_{n+1,1}&=\vvarr_{n,2},\\
    \vvarr_{n+1,2}&=\jacob_1^\supT\yvar_n+\vvarr_{n,3},\\
    \vvarr_{n+1,3}&=\vvarr_{n,4},\\
    \vvarr_{n+1,4}&=\jacob_2^\supT\yvar_n.
  \end{aligned}
\end{equation}
The segments of the compound delay line are not labeled here as in
Eqs.~\eqref{eq:coumpund_delay_line}, however, it is easy to see that
$\vvarr_{n,1}$ and $\vvarr_{n,2}$ belong to the first segment, and
$\vvarr_{n,3}$ and $\vvarr_{n,3}$ form the second one.

Altogether we have shown that the analytically derived adjoint variational
equation~\eqref{eq:main_adj_eq} agrees with the numerical one obtained
via straightforward computation of the adjoint matrix. In doing so,
the inner product~\eqref{eq:delay_inner_def} has to be used for
analytical treatments and Eq.~\eqref{eq:delay_inner_def_dis} is its
numerical vis-\'a-vis.

\section{\label{sec:num_scheme}Numerical procedure}

For actual computations, a numerical scheme based on the Euler
discretization is not the best choice due to its known poor accuracy.
We will employ the Heun's method to solve both the main
system~\eqref{eq:main_eq} and the variational
equation~\eqref{eq:main_var_eq}. This method belongs to the class of
second-order Runge-Kutta methods with constant time
step~\cite{NumericalDDE}. In the course of computations in addition to
the current state we have to keep also the data for $\ntau_\delnum$
previous steps along a trajectory to provide the retarded
variables. The advantage of the Heun's method for our problem is that
these stored values are enough for computations and no more additional
data are required for intermediate steps.

The Heun's step for the main system reads:
\begin{equation}
  \label{eq:heun_main_eq}
  \begin{aligned}
    \widetilde\xbas_{n+1}=&\xbas_n+h\fbas(t_n,\xbas_n,\xbas_{n-k_1},\ldots,\xbas_{n-k_\delnum}),\\
    \xbas_{n+1}=&\xbas_n+(h/2)\left[\fbas(t_n,\xbas_n,\xbas_{n-k_1},\ldots,\xbas_{n-k_\delnum})+
    \fbas(t_{n+1},\widetilde\xbas_{n+1},\xbas_{n+1-k_1},\ldots,\xbas_{n+1-k_\delnum})\right],
  \end{aligned}
\end{equation}
and the variational equation~\eqref{eq:main_var_eq} is solved as follows:
\begin{equation}
  \label{eq:heun_main_var_eq}
  \begin{aligned}
    \widetilde\xvar_{n+1}=&\xvar_n+h\left[\jacob_0(t_n)\xvar_n+\sum_{i=1}^{\delnum}\jacob_i(t_n)
      \xvar_{n-\ntau_i}\right],\\
    \xvar_{n+1}=&\xvar_n+h\left[\jacob_0(t_n)\xvar_n+\jacob_0(t_{n+1})\widetilde\xvar_{n+1}+
    \sum_{i=1}^{\delnum}\jacob_i(t_n) \xvar_{n-\ntau_i}+\jacob_i(t_{n+1}) \xvar_{n+1-\ntau_i}\right].
  \end{aligned}
\end{equation}

Though the backward time tangent space dynamics can be implemented via
straightforward solving Eq.~\eqref{eq:main_adj_eq}, it is more
efficient to reuse the data computed on the forward pass. Let us
define the following block matrix
$(\ntau_\delnum+1)\times(\ntau_\delnum+1)$ whose entries are zeros
except the first row:
\begin{equation}
  \label{eq:heun_matrix_rhs}
  \mat C(t_n)=
  \begin{pmatrix}
    \jacob_0(t_n) & c_1 & \dots & c_{\ntau_\delnum-1} & \jacob_\delnum(t_n)\\
    0 & 0 & \dots & 0 & 0\\
    \hdotsfor{5}\\
    0 & 0 & \dots & 0 & 0
  \end{pmatrix},
\end{equation}
where $c_i$ is either $\jacob_j(t_n)$ if $i$ coincides with one of the
delays $\ntau_j$ or zero: $c_i=\delta_{i,\{\ntau_j\}}\jacob_j(t_n)$.
Also we will need the matrix
\begin{equation}
  \mat S=
  \begin{pmatrix}
    1 & 0 & 0 & \dots & 0 & 0 \\
    1 & 0 & 0 & \dots & 0 & 0 \\
    0 & 1 & 0 & \dots & 0 & 0 \\
    0 & 0 & 1 & \dots & 0 & 0 \\
    \hdotsfor{6}\\
    0 & 0 & 0 & \dots & 1 & 0
  \end{pmatrix}.
\end{equation}
Using these matrices the forward time Heun's step can be represented
as $\bar\xvar_{n+1}=\numjac(t_n)\bar\xvar_{n}$, where
\begin{equation}
  \label{eq:heun_propagator}
  \numjac(t_n)=\frac{h}{2}\mat C(t_{n+1})[h\mat C(t_n)+\mat S]+
  \frac{h}{2}\mat C(t_n)+\mat S,
\end{equation}
and $\bar\xvar_n\in \mathbb{R}^{\locdim(\ntau_\delnum+1)}$ is a state
vector. We recall that $\locdim$ is a local dimension, i.e., the
dimension of a single vector variable $\xvar_n$.

The computations discussed so far required a non-standard inner
product Eq.~\eqref{eq:delay_inner_def_dis} with the
metric~\eqref{eq:the_metric}. But the traditional routines for linear
algebra manipulations in known numerical software libraries are
usually implemented with respect to the standard dot product. To bypass
this obstacle we can orthonormalize the tangent space basis using the
matrix $\hmatrix=\diagmat(1,\sqrt{h},\ldots ,\sqrt{h})$, see
Eq.~\eqref{eq:to_orthonorm_basis} and the related discussion. Thus
instead of a ``raw'' numerical Jacobian matrix $\numjac(t_n)$ the
modified one will be used that is defined with respect to the
orthonormal basis:
\begin{equation}
  \label{eq:modified_jac}
  \numjac'(t_n)=\hmatrix \, \numjac(t_n)\, \hmatrix^{-1}.
\end{equation}
If forward time tangent space
iterations~\eqref{eq:forward_lyap_step},~\eqref{eq:forward_qr} are
performed with $\numjac'(t_n)$, the adjoint matrix for the backward
time iterations~\eqref{eq:bkwadj_lyap_step},~\eqref{eq:bkwadj_qr} is
merely its transposition,
$[\numjac'(t_n)]^\adjnt=[\numjac'(t_n)]^\supT$, and the inner product
of the involved tangent vectors is computed via the standard dot
product.

Regardless of the high dimension of the phase space the numerical
Jacobian matrix contains sufficiently small number of nontrivial
trajectory dependent values. Its full size is
$\phdim^2=[\locdim(\ntau_\delnum+1)]^2$, but the non-constant values
are supplied only by the derivative matrices $\jacob_i(t)$, where
$i=0,1,\ldots,\ntau_\delnum$. The upper estimate for their total
number $(\delnum+1)\locdim^2$ is sufficiently small. But what is more
important, this number does not depend on the computation accuracy
that influences $\phdim$. It means that data for the numerical
Jacobian matrices can be stored along the trajectory without risk of
exhausting of a computer memory and then reused on the backward pass.

Thus the computations are organized as follows. The forward
steps~\eqref{eq:forward_lyap_step} are implemented via solving
Eqs.~\eqref{eq:heun_main_eq} and $\nang$ copies of
Eq.~\eqref{eq:heun_main_var_eq} over certain time intervals.
Solutions of the variational equations, treated as $(\ntau_\delnum+1)$
dimensional block vectors, each block of size $\locdim$, are
multiplied by $\hmatrix^{-1}$ before each step and by $\hmatrix$ after
the step.  It corresponds to the iteration with $\numjac'(t_n)$ in the
orthonormalized basis, see Eq.~\eqref{eq:modified_jac}. After the time
evolution step~\eqref{eq:forward_lyap_step} the vectors gathered as
columns of a matrix $\bkwpre(t)$ are QR-factorized as requires
Eq.~\eqref{eq:forward_qr}. The resulting matrices $\bkwmat(t)$
containing backward Lyapunov vectors are stored.  The nontrivial
values of the derivative matrices $\jacob_i(t)$ are also stored. The
backward pass~\eqref{eq:bkwadj_lyap_step},~\eqref{eq:bkwadj_qr} is
performed with
$[\numjac'(t_n)]^\supT=\hmatrix^{-1}[\numjac(t_n)]^\supT\hmatrix$,
where $\numjac(t_n)$ is recovered via Eq.~\eqref{eq:heun_propagator}
using the stored derivative matrices $\jacob_i(t)$. After each
QR-factorization~\eqref{eq:bkwadj_qr} the forward Lyapunov vectors
sitting in columns of $\fwdmat(t_n)$ are used together with the stored
backward Lyapunov vectors to compute the matrix $\mat P(t_n)$, see
Eq.~\eqref{eq:matrix_p}, and the angles $\theta_i$ as explained in
Sec.~\ref{sec:theory}.

\section{\label{sec:examples}Hyperbolicity testes of particular
  systems}

First we consider a generator of a robust chaos based on van der Pol
oscillator with two delayed feedbacks~\cite{Baranov10}:
\begin{equation}
  \label{eq:barkuzpon}
  \ddot\xbas-[A\cos(2\pi t/T)+B-\xbas^2]\dot\xbas+\omega_0^2\xbas=
  \epsilon\xbas(t-T/2)\,\xbas(t-3T/2)\dot\xbas(t-3T/2).
\end{equation}
Here $\xbas$ is a dynamical variable, $\epsilon$ controls the strength
of the delayed feedback and is supposed to be small; $A$ is the
amplitude of modulation of the excitation parameter with respect to the
middle level $B$. The period of modulation $T$ is assumed to be large
so that $T\gg 2\pi/\omega_0$, where $\omega_0$ is the natural
frequency of the oscillator. Note that this system was implemented as
a real electronic circuit and studied experimentally~\cite{Baranov10}.

The oscillator is activated and damped with the period $T$. The delay
times $T/2$ and $3T/2$ are selected in such way that every new activation stage is
initiated by signals from two previous subsequent activation stages. Suppose
these signals to be $\xbas\sim\sin(\omega_0 t+\phi_n)$ and
$\xbas\sim\sin(\omega_0 t+\phi_{n-1})$, respectively, i.e., their
phases are $\phi_n$ and $\phi_{n-1}$. Then, the nonlinear transformation
in the right hand side of Eq.~\eqref{eq:barkuzpon} provides a resonant
term:
\begin{equation}
  \xbas(t-T/2)\,\xbas(t-3T/2)\dot\xbas(t-3T/2)\sim
  (-1/4)\cos(\omega_0t-(5/2)\omega_0 T+2\phi_{n-1}-\phi_n)+\ldots,
\end{equation}
which stimulates the oscillation process arising at the new activation stage
imposing the own phase to it. It means that from stage to stage the
oscillation phase transforms according to the relation
\begin{equation}
  \label{eq:barkuzpon_map0}
  \phi_{n+1}=2\phi_{n-1}-\phi_n+\const,
\end{equation}
and, respectively, the phase difference $\Delta \phi_n=\phi_n-\phi_{n-1}$ obeys
the chaotic Bernoulli map
\begin{equation}
  \label{eq:barkuzpon_map}
  \Delta \phi_{n+1}=-2\Delta \phi_n+\const \mod 2\pi.
\end{equation}
The Bernoulli map~\eqref{eq:barkuzpon_map} is uniformly hyperbolic
with the Lyapunov exponent $\Lambda_1=\log 2\approx 0.693$. The
map~\eqref{eq:barkuzpon_map0} additionally has a zero Lyapunov exponent
$\Lambda_2=0$ related to its invariance with respect to an arbitrary
phase shift $\phi\to\phi+\alpha$.

Since the only mechanism responsible for chaos in the
system~\eqref{eq:barkuzpon} is uniformly hyperbolic, the stroboscopic
map for the system~\eqref{eq:barkuzpon} considered at $t_n=nT$ is also
expected to demonstrate the robust hyperbolic chaos. The detailed
analysis of its chaotic properties can be found in
Ref.~\cite{Baranov10}. Below we will verify its hyperbolicity using
the fast method of angles. For this and as well as for two other
systems discussed below, doing iterations~\eqref{eq:forward_lyap_step}
and~\eqref{eq:bkwadj_lyap_step} in the tangent space we will perform
QR-factorizations~\eqref{eq:forward_qr} and \eqref{eq:bkwadj_qr} at
each step of the discrete time.

Consider the system~\eqref{eq:barkuzpon} at two sets of the parameter
values:
\begin{subequations}
  \label{eq:barkuzpon_prm}
  \begin{alignat}{7}
    \label{eq:barkuzpon_prm_i}
    A=4, \;& B=0   \quad & (\Lambda=0.686, \;& 0.000, \;& -1.055, \;& -1.253, \;& \ldots);\\
    \label{eq:barkuzpon_prm_ii}
    A=5, \;& B=0.2 \quad & (\Lambda=0.668, \;& 0.000, \;& -1.990, \;& -2.118, \;& \ldots),
  \end{alignat}
\end{subequations}
where the other parameters are $T=8$, $\omega_0=2\pi$ and
$\epsilon=0.05$. In round brackets we specify four largest Lyapunov
exponents obtained numerically for the stroboscopic map of the
system~\eqref{eq:barkuzpon}. In agreement with the previous
discussion, the first Lyapunov exponents for both of the parameter
sets are close to $\log 2\approx 0.693$, and the second ones are zeros
within the numerical error.

\begin{figure}
  \onefig{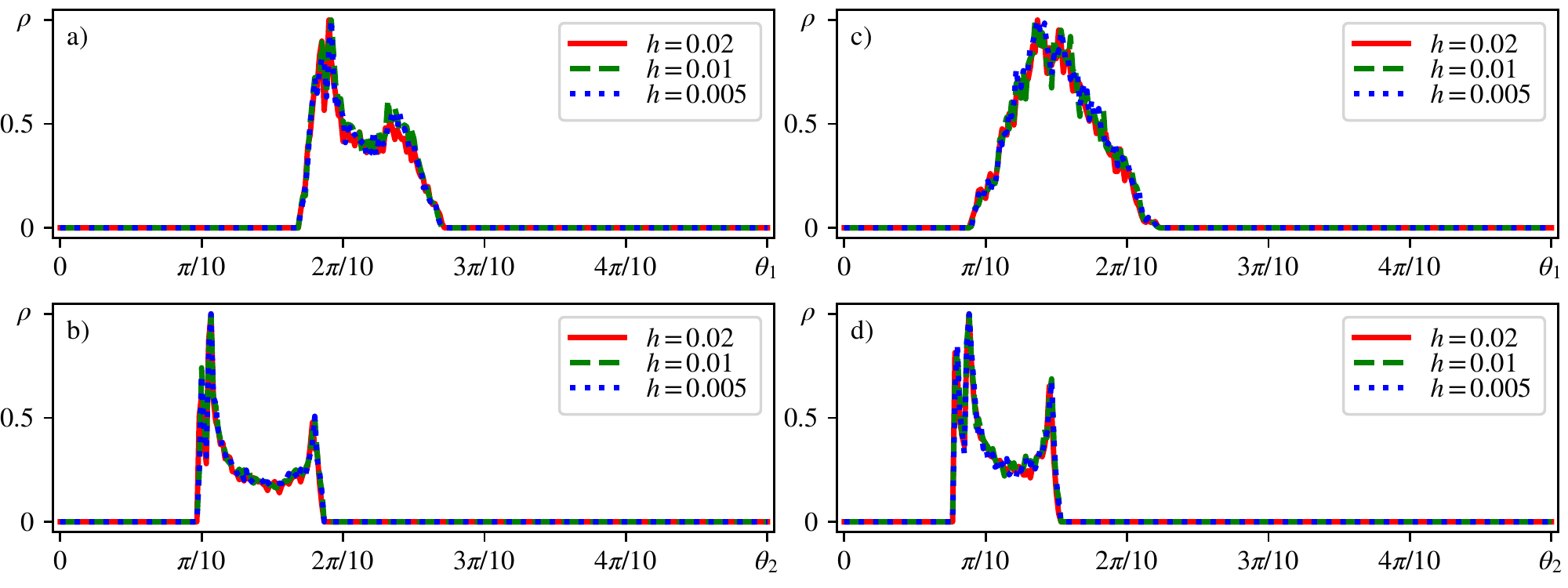}
  \caption{\label{fig:angdist_barkuzpon}(color online) Angle
    distributions for the system~\eqref{eq:barkuzpon} at different
    time steps $h$, see the legend. Panels (a) and (b) correspond to
    the parameter set~\eqref{eq:barkuzpon_prm_i}. In panel (a) the
    non-vanishing angle $\theta_1$ indicates that the expanding
    tangent subspace never touches the neutral plus contracting
    ones. The angle $\theta_2$ in panel (b) similarly indicates that
    sum of the expanding and neutral subspaces does not have
    tangencies with the contracting one. Together these two plots mean
    that the expanding, neutral and contracting subspaces are strictly
    separated from each other so that the system~\eqref{eq:barkuzpon}
    is hyperbolic. Panels (c) and (d) respectively confirm this
    conclusion for the parameter set~\eqref{eq:barkuzpon_prm_ii}.}
\end{figure}

Due to the presence of zero Lyapunov exponent, the tangent space of
the considered stroboscopic map splits into distinct subspaces:
expanding, neutral and contracting\footnote{Notice that unlike
  autonomous systems, where the neutral subspace is related to
  invariant time shifts and is eliminated with reformulation in terms
  of the Poincar\'e section map, in our case the neutral subspace
  appears due to the symmetry of the map itself.}.
Figure~\ref{fig:angdist_barkuzpon} provides the numerical conformation
that these three subspaces are disjoint.  This and all subsequent
figures have been plotted using Matplotlib graphics
package~\cite{Hunter:2007}. All diagrams are shown for three different
time steps $h$ to illustrate correspondence and convergence of the
computational data with the continuous limit.

Figs.~\ref{fig:angdist_barkuzpon}(a,b) correspond to the parameter
set~\eqref{eq:barkuzpon_prm_i}. Fig.~\ref{fig:angdist_barkuzpon}(a)
indicates that the distribution for angle $\theta_1$ is well separated
from the origin. It indicates that the expanding subspace never
clashes with the neutral and the contracting ones.
Figure~\ref{fig:angdist_barkuzpon}(b) indicates that the angle
$\theta_2$ is also separated from the origin; it means that the sum of
the expanding and neutral subspaces also has no tangencies with the
contracting subspace. Together Figs.~\ref{fig:angdist_barkuzpon}(a,b)
show that the expanding, neutral and contracting tangent subspaces
never touch each other, so that at any trajectory point the full
tangent space can be represented as their direct sum, that corresponds
to the main statement of the hyperbolicity concept.

The strict mathematical definition of the uniform hyperbolicity for
the discrete-time systems (diffeomorphisms) requires the existence of
the expanding and contracting subspaces only. Due to the presence of
the neutral subspace the stroboscopic map for the
system~\eqref{eq:barkuzpon} can be technically categorized as
partially hyperbolic~\cite{Pesin04}. However, the strict isolation of
the subspaces from each other indicates that the most important
property of the robustness of chaos nevertheless resides in this
system.

Data of testing for the second parameter
set~\eqref{eq:barkuzpon_prm_ii}, are plotted in
Figs.~\ref{fig:angdist_barkuzpon}(c,d). We also observe that the
angles are well separated from the origin that confirms the
hyperbolicity in this case too.

Let us now turn to another system with robust chaos introduced in
Ref.~\cite{KuzPik08} basing on an oscillator of the Stuart-Landau type
\begin{equation}
  \label{eq:qswdde}
  \dot a=(\gamma_0+\gamma_1\cos\Omega t-|a|^2)a+
  \epsilon [a(t-\tau)]^3 [a^*(t-\tau_1)]^2
\end{equation}
having in mind a possible implementation as a laser device.
Here $a$ is a complex dynamical variable; $\gamma_0$ and $\gamma_1$
control the excitation that is slowly modulated with the frequency
$\Omega$; $\tau$ and $\tau_1$ define the delay durations; small
$\epsilon$ controls the strength of the delayed feedback; asterisk
denotes the complex conjugation.

This system operates similarly to the previous one. It demonstrates
activation and damping with the period $T=2\pi/\Omega$. Since $\epsilon$
is small, the effect of the delayed signals is essential
at the beginnings of the activation stages.  Proper choice of the
delays provides transfer of the excitation for by the signals from two previous activation
stages. Being nonlinearly transformed, these signals produce a resonant
term whose phase depends on two previous phases as
\begin{equation}
  \label{eq:qswdde_map0}
  \phi_{n+1}=3\phi_n-2\phi_{n-1}+\const.
\end{equation}
Here $\phi_n$ is the phase of oscillations at the $n$th activation
stage. The phase difference $\Delta\phi_n=\phi_n-\phi_{n-1}$ evolve
according to the Bernoulli map:
\begin{equation}
  \label{eq:qswdde_map}
  \Delta \phi_{n+1}=2\Delta\phi_n+\const \mod 2\pi.
\end{equation}
This map has one positive Lyapunov exponent $\Lambda_1=\log 2$, and
the map \eqref{eq:qswdde_map0} additionally has the zero one due to a
symmetry related to an arbitrary phase shift. As studied in
Ref.~\cite{KuzPik08} the stroboscopic map of the system
\eqref{eq:qswdde} sliced at $t_n=nT$ demonstrates the robust chaos.

We consider the system~\eqref{eq:qswdde} stroboscopically for two
parameter sets:
\begin{subequations}
  \label{eq:qswdde_prm}
  \begin{alignat}{7}
    \label{eq:qswdde_prm_i}
    \gamma_0=0.2, \;& \gamma_1=2 \quad & (\Lambda=0.693, \;& 0.000, \;& -0.726, \;& -1.548, \;& \ldots);\\
    \label{eq:qswdde_prm_ii}
    \gamma_0=0.3, \;& \gamma_1=3 \quad & (\Lambda=0.693, \;& 0.000, \;& -1.174, \;& -1.172, \;& \ldots).
  \end{alignat}
\end{subequations}
Other parameters are $\Omega=1$, $\tau=5$, $\tau_1=11$,
$\epsilon=0.1$. The first Lyapunov exponents are close to
$\log 2$. The second exponents are zeros within the numerical error.

\begin{figure}
  \onefig{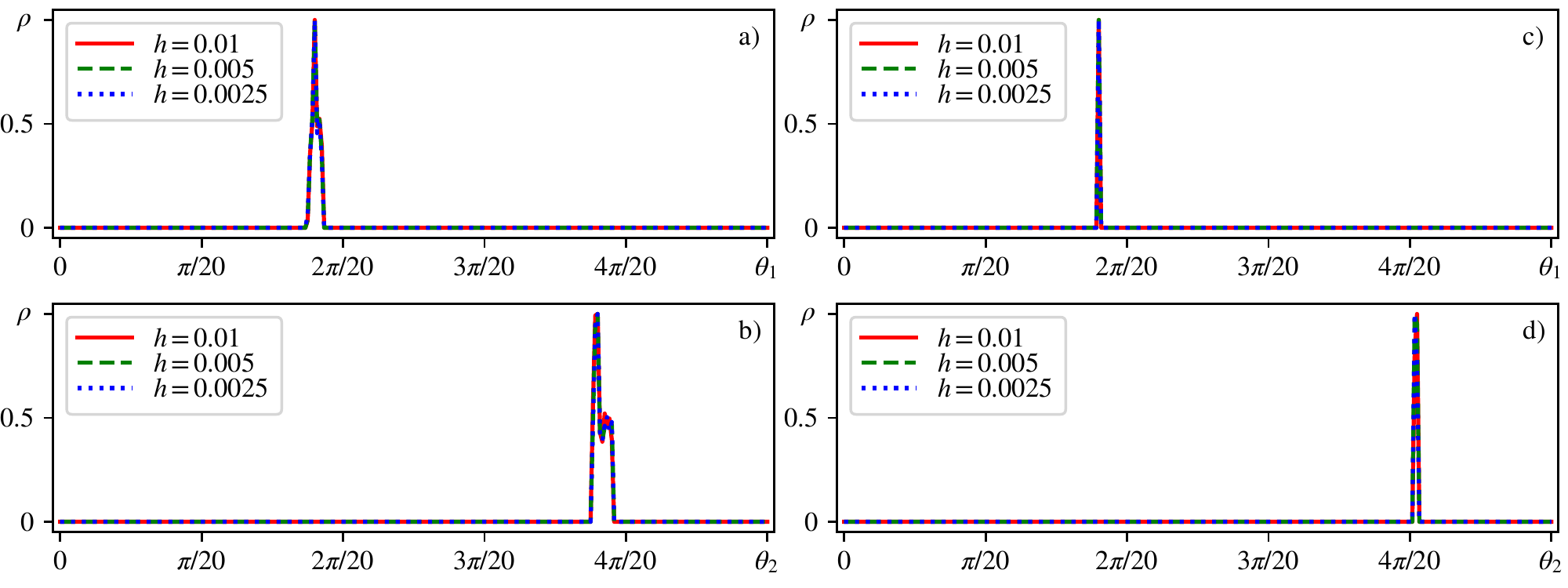}
  \caption{\label{fig:angdist_qswdde}(color online) Distributions of
    angles $\theta_1$ and $\theta_2$ for the system~\eqref{eq:qswdde}
    at different time steps $h$, see the legend. Panels (a) and (b)
    correspond to the parameter set~\eqref{eq:qswdde_prm_i}, and (c)
    and (d) represent the set~\eqref{eq:qswdde_prm_i}. Well defined
    separation of the angles from the origin confirms the
    hyperbolicity of the system~\eqref{eq:qswdde}.}
\end{figure}

Because of presence of a zero Lyapunov exponent, we again have to consider three
tangent subspaces: expanding, neutral and contracting ones.
Figures~\ref{fig:angdist_qswdde}(a,b) show the data of the computations for the parameter
set~\eqref{eq:qswdde_prm_i}. Both $\theta_1$ and $\theta_2$ are
well separated from the origin, so that we can conclude that the three
subspaces do not have tangencies. Analogously to the
system~\eqref{eq:barkuzpon}, it means that the stroboscopic map for
Eq.~\eqref{eq:qswdde} possesses the robust chaos and
can be categorized as partially hyperbolic~\cite{Pesin04}. Similar
analysis for the second parameter set~\eqref{eq:qswdde_prm_ii} also
confirms this conclusion.

Our final example is an autonomous generator of
robust chaos with two delays suggested in Ref.~\cite{Arzh14}:
\begin{equation}
  \label{eq:arzhkuz}
  \begin{aligned}
    \dot\xbas=&-\omega_0\ybas+(\mu/2)[1-\xbas^2(t-\tau_1)-\ybas^2(t-\tau_1)]\xbas+
    \epsilon[\xbas(t-\tau_1)\xbas(t-\tau_2)-\ybas(t-\tau_1)\ybas(t-\tau_2)],\\
    \dot\ybas=&\phantom{-}\omega_0\xbas+(\mu/2)[1-\xbas^2(t-\tau_1)-\ybas^2(t-\tau_1)]\ybas+
    \epsilon[\xbas(t-\tau_1)\ybas(t-\tau_2)+\xbas(t-\tau_2)\ybas(t-\tau_1)].
  \end{aligned}
\end{equation}
Here $\xbas$ and $\ybas$ are dynamical variables, $\mu$ is a
bifurcation parameter controlling the excitation, $\omega_0$ is a
natural frequency, and small $\epsilon$ determines the strength of the
delayed feedback.

The key idea of operation of this system is similar to the previous
ones: activation stages alternate with damping ones, and the delayed
signals provide the resonant seeds for every new excitation stage from
two preceding stages. In this case, however, the nonlinear
transformation of the delayed signals is of such kind that evolution
of the phases from stage to stage take place according to chaotic
Anosov torus map. One more difference is that this system is
autonomous: alternation of the activation and damping stages occurs
due to the internal dynamics of the system, without external
modulation of parameters.

Chaotic properties of Eq.~\eqref{eq:arzhkuz} are studied in detail in
Ref.~\cite{Arzh14}. In particular it is shown that new phase at
$(n+1)$th stage of excitation $\phi_{n+1}$ depends on two previous
phases according to the Fibonacci map
\begin{equation}
  \label{eq:arzhkuz_map}
  \phi_{n+1}=\phi_n+\phi_{n-1}+\const \mod 2\pi.
\end{equation}
This map has two Lyapunov exponents that are equal to logarithms of
golden ratio and reciprocal golden ratio respectively:
$\Lambda=\pm\log [(1+\sqrt{5})/2]\approx \pm 0.481$.

To check hyperbolicity of the system~\eqref{eq:arzhkuz} we construct a
Poincar\'e section map whose iterations correspond to the successive
excitation stages. We define this map using the section surface in the
state space determined by the relation
\begin{equation}
  \label{eq:arzhkuz_section}
  \xbas^2+\ybas^2=1.
\end{equation}

Consider two sets of the parameter values:
\begin{subequations}
  \label{eq:arzhkuz_prm}
  \begin{alignat}{7}
    \label{eq:arzhkuz_prm_i}
    \mu=1.6, \;& \epsilon=0.02 \quad & (\Lambda=0.481, \;& 0.000, \;& -0.473, \;& -0.530, \;& \ldots);\\
    \label{eq:arzhkuz_prm_ii}
    \mu=2,   \;& \epsilon=0.05 \quad & (\Lambda=0.481, \;& 0.000, \;& -0.013, \;&  -0.481, \;& \ldots).
  \end{alignat}
\end{subequations}
The other parameters are $\omega_0=2\pi$, $\tau_1=2$, $\tau_2=14$. In
round brackets we specify four largest Lyapunov exponents obtained
numerically for the corresponding Poincar\'e map of the system.
Observe that the largest Lyapunov exponents in both cases correspond
as expected to logarithms of the golden ratio. Moreover, among the
negative exponents one is equal approximately to the logarithm of the
reciprocal golden ratio ($-0.473$ for the first case and $-0.481$ for
the second one). The second Lyapunov exponents are zero, as it is
typical for autonomous flow systems. (Though we consider the
Poincar\'e map, i.e., a discrete time system, the zero exponent still
exists since in the course of computations we evaluate the Lyapunov
vectors within the full tangent space. To eliminate zero we could find
projection of the Lyapunov vectors onto the section surface, however,
this it is not needed in the context of our consideration.)

For comparison, we also have computed the first four Lyapunov
exponents for the original flow system~\eqref{eq:arzhkuz}:
\begin{subequations}
  \label{eq:arzhkuz_lam}
  \begin{alignat}{5}
    \label{eq:arzhkuz_lam_i}
    \lambda=0.044, \;& 0.000, \;& -0.043, \;& -0.050, \;& \ldots \\
    \label{eq:arzhkuz_lam_ii}
    \lambda=0.053, \;& 0.000, \;& -0.002, \;& -0.053, \;& \ldots
  \end{alignat}
\end{subequations}
Rows~\eqref{eq:arzhkuz_lam_i} and~\eqref{eq:arzhkuz_lam_ii} correspond
to parameters~\eqref{eq:arzhkuz_prm_i} and~\eqref{eq:arzhkuz_prm_ii},
respectively. Observe that the largest exponents do not coincide since
the periods of excitation in the two cases are different. If we
considered the trajectory and found the average return periods
$\langle T\rangle$, we would obtain that
$\Lambda_1=\langle T\rangle\lambda_1$ for both cases, in agreement
with the computations for the Poincar\'e map.

Let us now turn to the hyperbolicity check. The initial system is
autonomous with a single zero Lyapunov exponent related to the
symmetry with respect to shifts in continuous time. Introducing the Poincar\'e
section we exclude this symmetry, so the zero exponent disappears. It
means that the hyperbolicity test in this case must include
computation of one angle only, between the expanding and contracting
tangent subspaces. However as already mentioned, it would require
computation of projections of Lyapunov vectors onto the section
surface that is computationally inefficient. Instead we will check if
the system~\eqref{eq:arzhkuz} corresponds to the Anosov
flow~\cite{Anosov95,KatHas95} on the section surface. It automatically
implies the hyperbolicity of the corresponding Poincar\'e map.

\begin{figure}
  \onefig{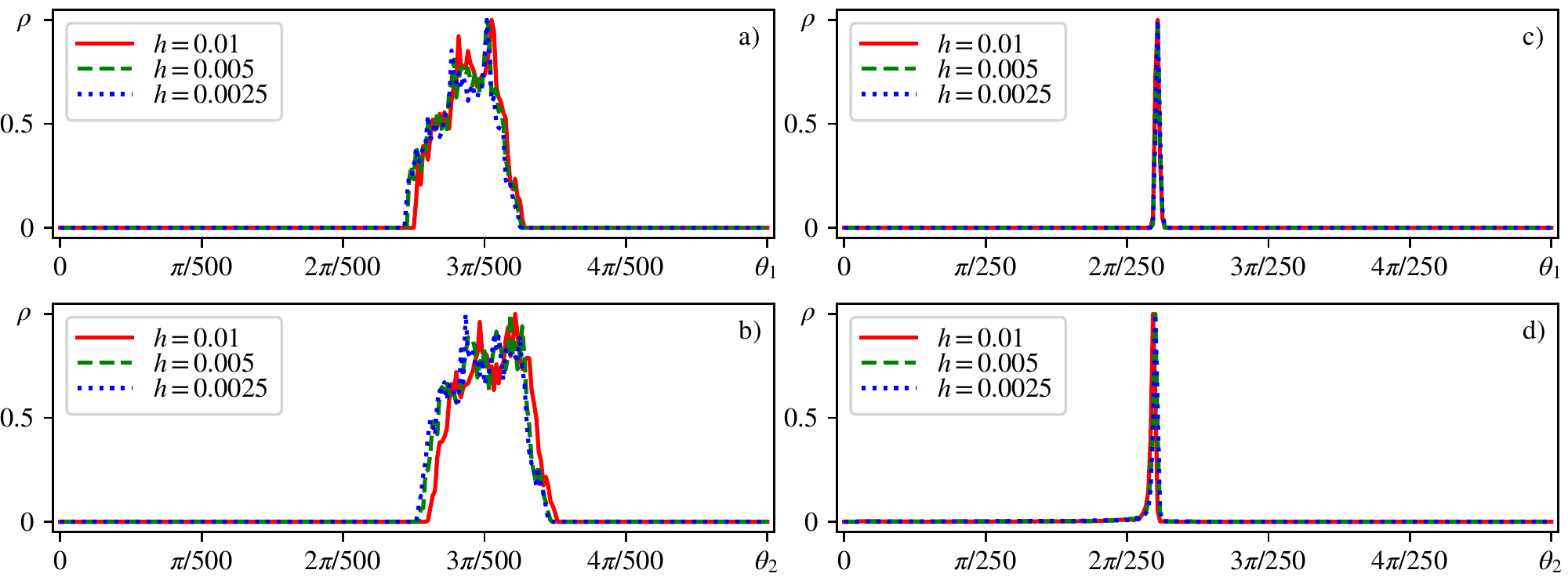}
  \caption{\label{fig:angdist_arzhkuz}(color online) Distributions of
    angles $\theta_1$ and $\theta_2$ at different time steps $h$, see
    the legend, for the system~\eqref{eq:arzhkuz} on the section
    surface~\eqref{eq:arzhkuz_section}. Panels (a,b) and (c,d)
    corresponds to parameter sets~\eqref{eq:arzhkuz_prm_i}
    and~\eqref{eq:arzhkuz_prm_ii}, respectively. Well separated from
    the origin distributions confirm the hyperbolicity in both cases.}
\end{figure}

For this purpose we again obtain numerically the distribution of the
angles $\theta_1$ for the expanding vs. the neutral plus contracting
subspaces, and the distribution of the angles $\theta_2$ for the
expanding plus neutral vs. the contacting
subspaces. Figures~\ref{fig:angdist_arzhkuz}(a,b) and (c,d) show that
the system indeed is hyperbolic in both cases~\eqref{eq:arzhkuz_prm_i}
and~\eqref{eq:arzhkuz_prm_ii}.

\section{\label{sec:concl}Summary}
Systems with hyperbolic chaotic attractors including those with
time-delays are of great potential importance for applications because
of the intrinsic structural stability that implies nonsensitivity of
the generated chaos to parameters and functional characteristics of
the components, to perturbations, noises, interferences, fabrication
imperfections and so on. In this paper a method of computer
verification of hyperbolic nature of chaotic attractors is developed
in a form appropriate for systems with multiple time delays. The
method is based on the calculation of the angles between expanding,
contracting and neutral manifolds for phase trajectories (the ``angle
criterion'').

Since for time-delay systems the phase space is infinite-dimensional,
the contracting manifold is also infinite-dimensional. Therefore, the
procedure is based on the use of the complement to a contracting
tangent subspace. For tangent vectors related to the expanding and
neutral subspaces linearized equations with retarded argument are
integrated along a reference trajectory on the attractor in direct
time. The contracting subspace is identified with the use of the
adjoint system of equations with deviating argument of leading or
advanced type, formulated within the framework of a specially worked
out mathematical justification of the technique. The integration of
the adjoint equations is performed along the reference trajectory in
the inverse time. The obtained data make it possible to obtain and
analyze statistics for distributions of the angles of intersection of
the expanding, contracting and neutral subspaces in the tangent space
of deviation vectors for the reference trajectory. The absence of
angles close to zero indicates hyperbolicity, while a nonzero
probability of zero angles implies its violation. With the help of the
proposed algorithm, the hyperbolic nature of chaos is substantiated
for three examples of time-delay systems with two delays by
presentation of histograms of angular distributions.

\bigskip
\textit{Work of SPK on theoretical formulations was supported by grant of
Russian Science Foundation No 15-12-20035. The work of PVK on
elaborating computer routines and numerical computations was supported
by grant of RFBR No~16-02-00135.}
\bigskip

\bibliography{hyp_many_delay}

\end{document}